\documentclass{emulateapj}

\usepackage{amsmath}
\usepackage{array}
\usepackage{bm}
\usepackage{cases}
\usepackage{txfonts}
\usepackage{tipa}
\usepackage{amsfonts}
\usepackage{amssymb}
\usepackage{graphicx}
\usepackage{multirow}
\usepackage{float}
\usepackage{longtable}

\shortauthors{Du Cuihua et al.}

\begin{document}

\title{The origin of high velocity stars from Gaia and LAMOST}

\author{Cuihua Du\altaffilmark{1}, Hefan Li\altaffilmark{2}, Heidi Jo Newberg\altaffilmark{3},Yuqin Chen\altaffilmark{4,1}, Jianrong Shi\altaffilmark{4,1},Zhenyu Wu\altaffilmark{4,1}, Jun Ma\altaffilmark{4,1}}

\affil{$^{1}$College of Astronomy and Space Sciences, University of Chinese Academy of Sciences, Beijing 100049, China; ducuihua@ucas.ac.cn\\
$^{2}$School of Physical Sciences, University of Chinese Academy of Sciences, Beijing 100049,  China \\
$^{3}$Department of Physics, Applied Physics and Astronomy, Rensselaer Polytechnic Institute, Troy, NY 12180, USA, newbeh@rpi.edu\\
$^{4}$Key Laboratory of Optical Astronomy, National Astronomical Observatories, Chinese Academy of Sciences, Beijing 100012, China\\
}

\begin{abstract}
\par Based on the second Gaia data (Gaia DR2) and spectroscopy from the LAMOST Data Release 5, we defined the high-velocity (HiVel) stars sample as those stars with $v_{\mathrm{gc}} > 0.85 v_{\mathrm{esc}}$, and derived the final sample of 24 HiVel stars with stellar astrometric parameters and radial velocities. Most of the HiVel stars are metal-poor and $\alpha$-enhanced. In order to further explore the origin of these HiVel stars, we traced the backwards orbits of each HiVel star in the Galactic potential to derive probability parameters which are used to classify these HiVel stars. Of these, 5 stars are from the tidal debris of disrupted dwarf galaxy and 19 stars are runaway-star candidates which originate from the stellar disk.
\end{abstract}


\section{Introduction}

\par High-velocity (HiVel) stars move sufficiently fast so that they could escape the gravitational potential of the Galaxy. With the development of large spectroscopic surveys such as SDSS, RAVE, LAMOST and Gaia, a large number of high velocity candidates have been reported \citep[e.g.,][]{Brown06, Brown09, Brown12, Brown14, Li12, Li18, Zheng14,  Zhong14, Geier15, Zhang16, Huang17, Du18, Marchetti18, Bromley18} .  HiVel stars are intriguing because they not only flag the presence of extreme dynamical and astrophysics processes, but also can be used as dynamical traces of integral properties of the Galaxy. In particular,  the origin of HiVel stars can provide 
useful information about the environments from which they are produced.  In general, there are three subclasses for HiVel stars and they have different origin.  First of all, the fastest stars in our Galaxy are hypervelocity stars (HVSs),  which have extreme velocities above the escape speed of  the Milky Way.  
HVSs can obtain their large velocity from a number of different processes.  \citet{Hills88}  first theoretically predicted the formation of HVSs via three-body interactions between a binary star system and the massive black hole (MBH) in the Galactic Center (GC).  Other possible alternative mechanisms also include the interaction  between single stars and a hypothetic binary MBH \citep{Yu03, Sesana06,Sesana07,Merritt06}, the interaction between a globular cluster with a single or a binary MBH in the GC \citep[e.g.,][]{Capuzzo15, Fragione16}. Since the first HVS was discovered by \cite{Brown05},  almost two dozen unbound HVSs of late B type with masses between 2.5 and 4 M$_{\odot}$ \citep{Brown14, Zheng14, Geier15, Huang17} have been found from systematic searches.  Besides the unbound population of HVSs ,  all mechanisms mentioned above  also predicted a population of bound HVSs \citep{Bromley09}.   For example, \citet{Brown14} identified 16 such stars whose Galactic rest-frame velocities exceed 275 kms$^{-1}$.  

\par ``Runaway stars" are another subclass of high velocity stars and were first introduced as O and B type stars by \citet{Blaauw61}.  Runaway stars are thought to have formed in the disk and ejected into the halo. These stars can provide important connection between star formation 
in the Galactic disk and halo.  In general, runaway  stars can be produced through two main formation mechanisms, i.e. (1) supernova explosions in stellar binary systems \citep[e.g.][]
{Blaauw61,Portegies00,Gvaramadze09,Wang13} and (2) dynamical encounters due to multi-body encounter in dense stellar systems \citep[e.g.,][]{Bromley09, Gvaramadze09}. Both mechanisms can produce both low-mass and high-mass runaway stars. But majority of runaway stars in the literature are high-mass O and B type stars with ejection velocities less than 200 kms$^{-1}$ \citep{Perets12}.  Recent results show it is possible for low-mass G/K type stars with ejection velocities up to $\sim1300$ kms$^{-1}$\citep{Tauris15}. Besides the two classes HiVel stars mentioned above,  there also exists fast halo stars from the tidal debris of an accreted and disrupted dwarf galaxy \citep{Abadi09, Teyssier09}. 

\par In order to distinguish between the scenarios, recent studies have used the chemical and kinematic information to determine the origin of HiVel stars \citep[e.g.,][]{Hawkins15, Li12, Geier15,Marchetti18}. 
For example,  if HiVel stars are more metal-rich ([Fe/H]$>-0.5$) than expected for the inner halo, and the [$\alpha$/Fe] measurements are consistent with those of disk stars, it may suggest  that these metal-rich HiVel stars formed in the disk and were subsequently dynamically ejected into the halo \citep{Bromley09, Purcell10,Hawkins15}.  The kinematic studies need to use accurate proper motions and parallaxes to calculate trajectories with sufficiently small uncertainties. 
The second Gaia data release of Gaia survey \citep{Gaia18} provide an unprecedented sample of precisely and accurately measured source. 

\par In this letter, we use Gaia proper motions \citep{Gaia16a, Gaia16b} and radial velocities combined with radial velocities and metallicities derived from LAMOST stellar spectra \citep{Zhao12} to study the origin of HiVel stars. In Section 2, we briefly describe the data and target selection. In Section 3, we identify these  HiVel stars and explore their origin, including an analysis of the chemical abundances and orbital properties. The conclusions and summary are given in Section 4.

\section{Data and target selection}

\subsection{Data}

\par The second Gaia data (Gaia DR2) includes high-precision measurements of nearly 1.7 billion stars \citep{Gaia18}. As well as positions, the data include astrometry, photometry, radial velocities, and information on astrophysical parameters and variability, for sources brighter than magnitude 21. This data set contains parallaxes, and mean proper motions for about 1.3 billion of the brightest stars. Radial velocity measurements $rv_{G}$ for a subset of 7,224,631 stars are included in the Gaia DR2 with an effective temperature from 3550 to 6990 K, and the typical uncertainties are a few hundreds of m/s at the bright end of Gaia G magnitude and, a few km/s at the faint end. 
In the following we will focus on the subsample of stars.

\par The Large Sky Area Multi-object Fiber Spectroscopic Telescope (LAMOST) is a 4 meter quasi-meridian reflective Schmidt telescope, wich is equipped with 4000 fibers within a field of view of $5^{\circ}$. The LAMOST spectrograph has a resolution of R $\rm \sim$ 1,800 and wavelength range spanning 3,700 {\AA} to 9,000 {\AA} \citep{Cui12}.The survey reaches a limiting magnitude of $r=17.8$ (where $r$ denotes magnitude in the SDSS $r$-band), but most targets are brighter than $r\sim17$. The LAMOST Stellar Parameter Pipeline \citep{Wu11,Luo15} estimates parameters, including radial velocity, effective temperature, surface gravity and metallicity ([Fe/H]) from LAMOST spectra. The accuracies in measuring radial velocity ($rv_L$) and [Fe/H] at R = 1800 are expected to be 7 km/s and 0.1 dex, respectively \citep{Deng12, Zhao12}. The LAMOST Stellar Parameter Pipeline at Peking University [LSP3] \citep{Xiang15, Xiang17} gives $\alpha$-element to iron abundance ratio [$\alpha$/Fe]. In total, there are over 5 million stars in the A, F, G and K type star catalog. 

\par From the quasars and validation solutions, \citet{Lindegren18} estimated that systematics in the parallaxes depending on position, magnitude, and color are generally below 0.1 mas, but the parallaxes are on the whole too small by about 0.029 mas. The radial velocity zero-points (RVZPs) of large-scale stellar spectroscopic surveys need to be determined and corrected for further studies. \citet{Huang18} presented a new catalogue of 18,080 radial velocity standard stars selected from the APOGEE data. To determine the RVZP of LAMOST measurements, we cross-match the APOGEE radial velocity (RV) standard stars with the LAMOST DR5 catalogue and obtain 3,580 common stars of LAMOST spectral SNR greater than 20. The stars yield a mean difference $\Delta$rv = $-4.70$ km/s and a standard deviation s.d. = 4.45 km/s. We also cross-match the APOGEE RV standard stars with Gaia DR2 and obtain 8,786 common stars. The mean difference found by these stars is $\Delta$rv = 0.47 km/s, with a standard deviation s.d. = 1.40 km/s. We calibrate the parallax and radial velocity measurements with determined offsets in the following study.

\subsection{HiVel Candidate Selection}
\label{sample}

\begin{figure}[b]
	\includegraphics[width=1.0\hsize]{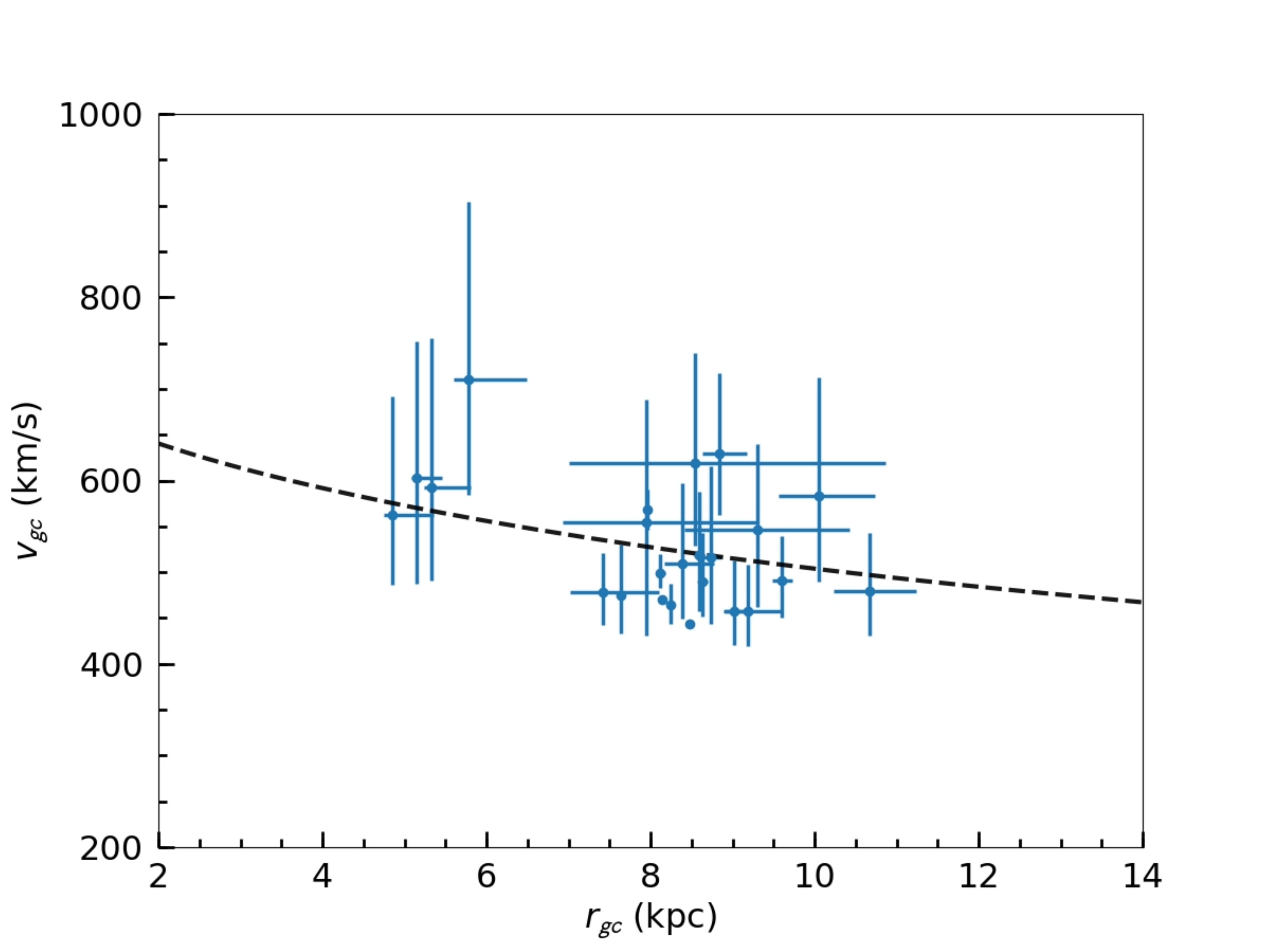}
	\caption{Total velocity in the Galactic rest-frame $v_{\mathrm{gc}}$ as a function of Galactocentric distance $r_{\mathrm{gc}}$ for 24 HiVel stars. The black dashed line is the median escape speed from \citet{Williams17} and the blue dot represent the HiVel stars sample.}
	\label{figure1} 
\end{figure}

\par Our initial sample was obtained by cross-matching between the Gaia and LAMOST catalogs based on stellar position.  
We first select those stars with signal-to-noise ratio (SNR) $\geq 20$. In order to ensure the reliable radial velocity, we also use the selection criterion $|rv_G - rv_L| \leq 10$ km/s. We adopt the weighted means for radial velocity and its error:
\begin{equation*}
rv = \frac{rv_G\,\sigma_L^2 + rv_L\,\sigma_G^2}{\sigma_G^2 + \sigma_L^2},\ \sigma_{rv}^2 = \frac{\sigma_L^2\,\sigma_G^2}{\sigma_G^2 + \sigma_L^2}
\end{equation*}
where $G$ and $L$ represent Gaia and  LAMOST.

\par Adopting the method from \citet{Luri18}, we use Bayesian analysis to determine the distance and velocity of the stars. We adopt the exponentially decreasing space density prior in distance $d$ \citep{Bailer18}:
\begin{equation*}
P(d\ |\ L) \propto d^2 \exp (-d/L)
\end{equation*}
and assume uniform priors on $v_{ra}, v_{dec}, v_r$. So we can express the posterior distribution:
\begin{equation*}
P(\bm{\theta}\ |\ \bm{x}) \propto \exp [-\frac{1}{2} (\bm{x} - \bm{m(\theta)})^\mathrm{T} C_x^{-1} (\bm{x} - \bm{m(\theta)})]\ P(d\ |\ L)
\end{equation*}
where $\bm{\theta} = (d,\ v_{ra},\ v_{dec},\ v_r)^\mathrm{T}$, $\bm{x} = (\varpi,\ \mu_{\alpha^*},\ \mu_{\delta},\ rv)^\mathrm{T}$, $\bm{m} = (1/d,\ v_{ra}/kd, \ v_{dec}/kd,\ v_r)^\mathrm{T}$, $k$ = 4.74 and $C_x^{-1}$ is covariance matrix. The positions and velocities are derived from the most probable value of $d, v_{ra}, v_{dec}, v_r$.

Total velocities in the Galactic rest frame are computed correcting radial velocities and proper motions for the solar and the local standard of rest motion. 
Here, the distance of the Sun from the Galactic center $R_{\odot}$ = 8.2 kpc, and the Sun has an offset from the local disk $z_{\odot}$ = 25 pc \citep{Bland16}.
We calculate each star's Galactic space-velocity components, $U$, $V$ and $W$, from its tangential velocities, distance, and radial velocity \citep{Johnson87}. 
We assume the LSR velocity is $V_\mathrm{{LSR}}$ = 232.8 km/s in the direction of rotation \citep{McMillan17} and the solar peculiar motion $(U, V, W) = (10., 11., 7.)$ km/s \citep{Tian15, Bland16} relative to the local standard of rest (LSR). The median escape speed $v_{\mathrm{esc}}$ can be derived from \citet{Williams17}. Applying the criterions mentioned above and further constrain on the total velocity $v_{\mathrm{gc}} > 0.85 v_{\mathrm{esc}}$, we obtain 37 candidates of HiVel stars.

\begin{figure*}[t]
	\includegraphics[width=1.0\hsize]{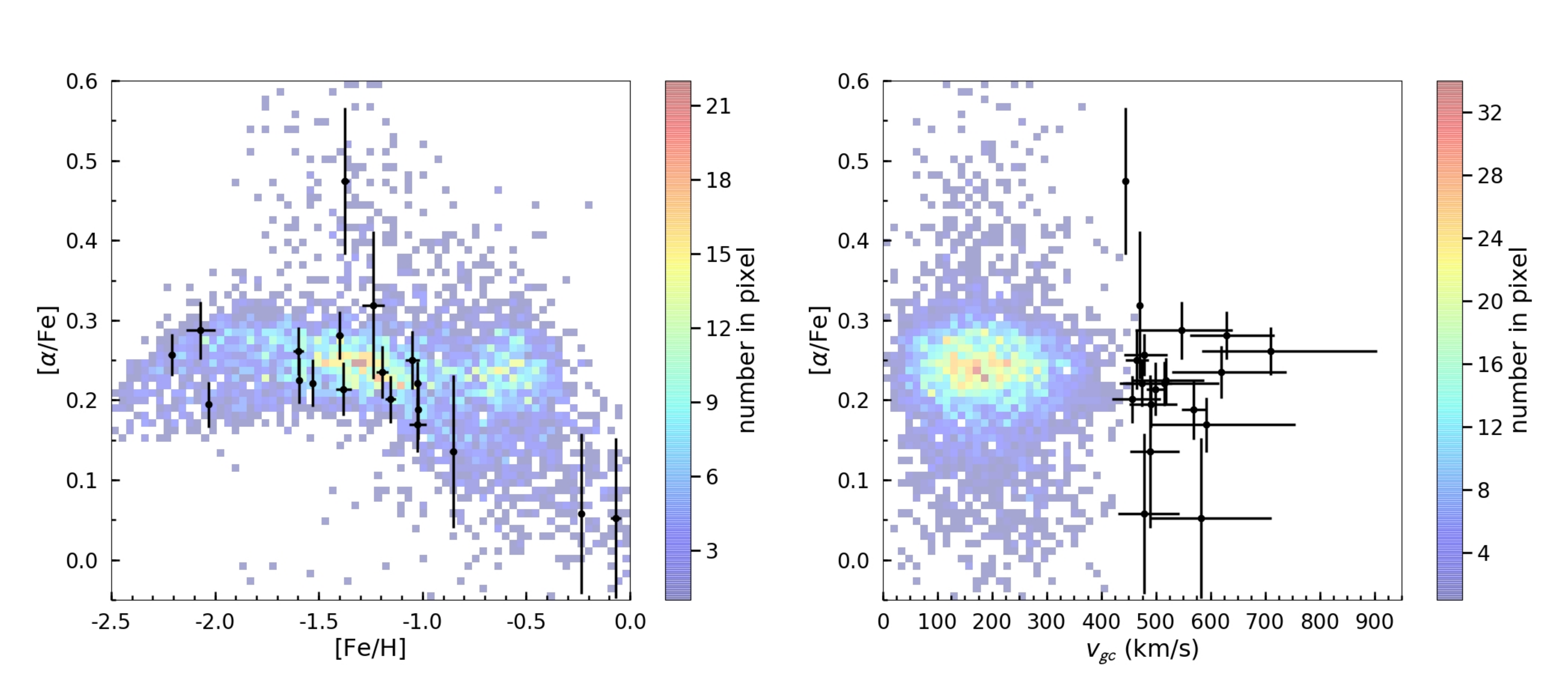} 
	\caption{Left panel: Chemical abundance distribution [$\alpha$/Fe] vs. [Fe/H] of 24 HiVel stars. Right panel: Chemical abundance distribution [$\alpha$/Fe] vs. $v_{\mathrm{gc}}$.  The black points represent the HiVel stars. The halo stars that selected by Toomre Diagram are shown as background for comparison and the color coding corresponds to the number of halo stars in each pixel.}
	\label{figure2}
\end{figure*} 

Then we use Markov chain Monte Carlo (MCMC) sampler $\textsc{EMCEE}$ to estimate error of these stars. We use 20 walkers and sample for 200 iterations. We run 1000 burn-in steps to let the walkers find starting point. In order to filter out the uncertain candidates, we remove stars with $\sigma_{\mathrm{vgc}}\,/v_{\mathrm{gc}} < 0.3$ and $\sigma_{\mathrm{rgc}} < 2$ kpc. Finally, we get 24 HiVel stars sample.

Figure \ref{figure1} shows the total velocity in the Galactic rest-frame $v_{\mathrm{gc}}$ as a function of Galactocentric distance $r_{\mathrm{gc}}$ for 24 HiVel stars. Most of our high velocity stars lie in the inner region of the Galaxy. The catalog of 24 HiVel stars is given in the Appendix. 

\section{Chemical abundances and orbits of HiVel stars}

\par The distribution in [$\alpha$/Fe] space also provides valuable information on the timescales and intensities of star formation in the populations involved. The study by \cite{Nissen10}  proposed that the high-$\alpha$ stars may have been born in the disk or bulge of the Milky way and heated to halo kinematics by merging satellite galaxies or else were simply members of the early generations of halo stars born during the collapse of a proto-Galactic gas cloud, while the low-$\alpha$  stars may have been accreted from dwarf galaxies. Therefore, the abundance space of [$\alpha$/Fe] versus [Fe/H] is particularly useful in tracing the origin of individual stars \citep{Lee15}. Figure \ref{figure2} shows the chemical abundance distribution [$\alpha$/Fe]  versus [Fe/H] and [$\alpha$/Fe] versus $v_{gc}$ for some HiVel stars. For comparison, we also add the halo stars as background in the figure.  The halo stars are defined as having $|v_{gc}-v_{\textrm{LSR}}| > 232.8 $ km/s, where $v_{\textrm{LSR}}= (0, 232.8, 0)$ km/s in the Galactocentric Cartesian coordinates.  
We can see from Figure \ref{figure2} that most of our HiVel stars are metal-poor and slightly $\alpha$-enriched, with a mean $\alpha$-abundance of $\overline{[\alpha/\rm Fe]}=+0.22$ dex, which is consistent with the result of \cite{Hawkins15}, with a mean $\alpha$-abundance of $\overline{[\alpha/ \rm Fe]}=+0.24$ dex. 
It shows some HiVel stars could have originated from the Galactic center or disk, while some from dwarf galaxies. For example, GLHV-8 has high [Fe/H] = $-$0.24 and low [$\alpha$/Fe] = 0.06. It seems like coming from thick disk.
The large dispersion in the [$\alpha$/Fe] could result from the uncertainty of the individual [$\alpha$/Fe] estimates. The large uncertainty in the [$\alpha$/Fe] estimates, is a result of the relatively low resolution of LAMOST spectra. We are looking forward to high resolution spectra of these stars in the future.

\begin{figure*}
	\centering
\includegraphics[width=1.0\vsize]{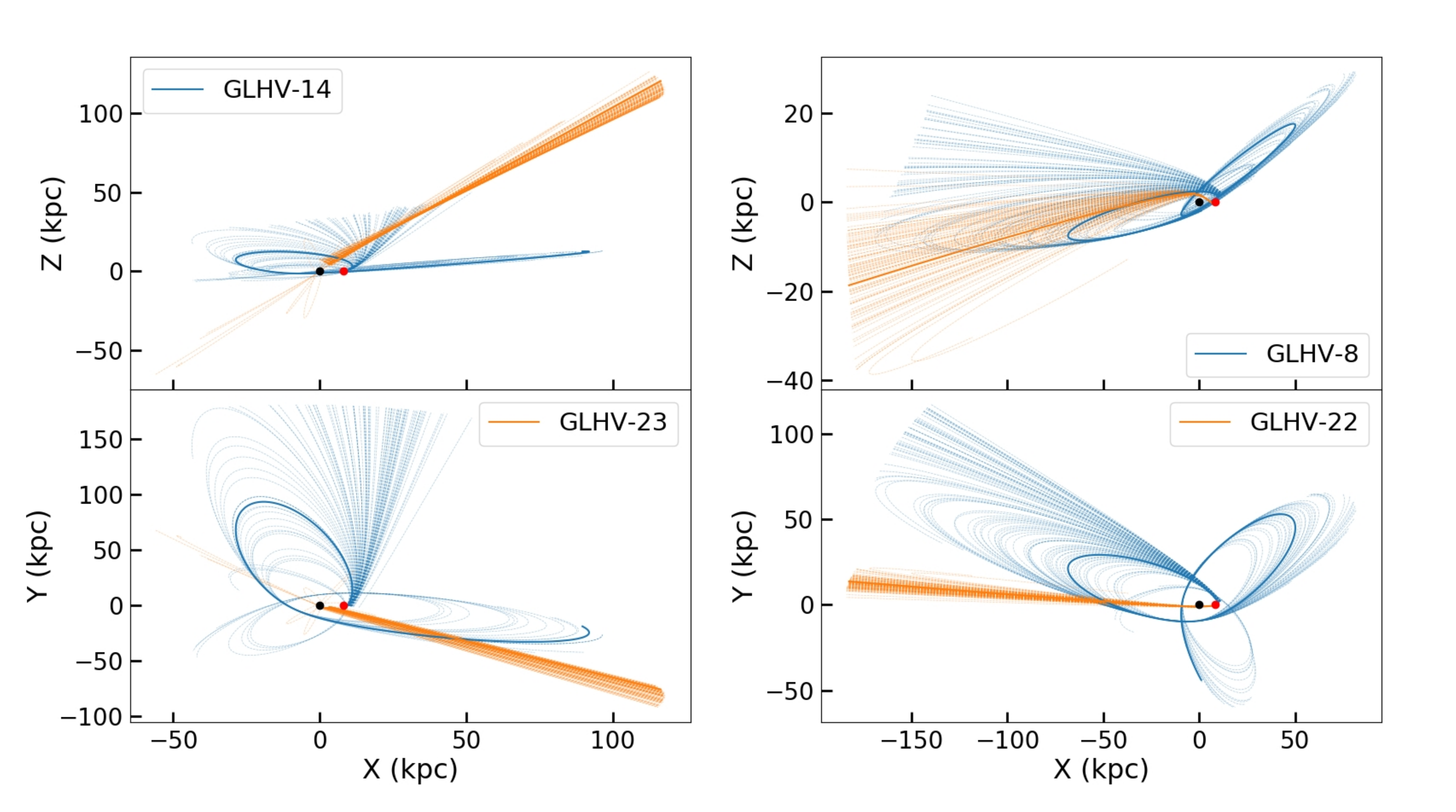} 
\caption{2 Gyr backwards orbit of the some represented HiVel stars in $XYZ$ Galactocentric Coordinates. 
The red dot represents the Sun and the black dot represents the Galactic Center. The thin lines show 100 orbits drawn at random from the uncertainties in the positions and velocities of each HiVel star, showing the uncertainty in the orbits.}
\label{figure3}
\end{figure*}

\par To get some hints on the ejection location of our HiVel stars,  we study their orbital properties by adopting a Galaxy potential model provided in \citet{McMillan17}. 
This model includes components that represent the contribution of the cold gas discs near the Galactic plane, as well as thin and thick stellar discs, a bulge component and a dark-matter halo. For each star, we use 4000 MCMC realizations discussed in Sect. \ref{sample}. We integrate each orbit back in a total time of 2 Gyr, starting with the current position of each star. If a star reaches the maximum potential point $\Phi_{\mathrm{max}}$ in 2 Gyr, we will call it unbound. The orbit will be cut off at that point to ensure its reliability and we can get the probability of a star being unbound $P_{\mathrm{ub}}$.

As an example, Figure \ref{figure3} gives the derived backward orbits for three subclasses HiVel stars, integrated back 2 Gyr. The red dot represents the present position, and the black dot represents the Galactic Center. 

\begin{center}
	\begin{longtable}{cccc}
	      \caption{The probability of stars are used as the classified criteria of HiVel stars}\\
		\hline
		\hline
		Class & $P_{\mathrm{gc}}$ &   $P_{\mathrm{MW}}$   &   $P_{\mathrm{ub}}$  \\
		\hline
		HVS	candidates & $>0.16$ & - 	   & - \\
		OUT	candidates& $<0.16$ & $>0.5$ & - \\
		HRS	candidates & $<0.16$ & $<0.5$ & $>0.5$ \\
		RS  candidates& $<0.16$ & $<0.5$ & $<0.5$ \\
		\hline
	\end{longtable}
\label{class}
\end{center}

Adopting the method from \citet{Marchetti18}, we could derive the position of a star crossing the disk and calculate the distance from the Galactic center to the crossing point. The minimum value of the distance is called $R_{\mathrm{min,\,2Gyr}}$. Some stars' velocities are slightly smaller than $v_{\mathrm{esc}}$, maybe 2 Gyr could not be enough for them to cross the disk. So we increase the trace-back time to 5 Gyr and get the minimum crossing radius $R_{\mathrm{min,\,5Gyr}}$ just like above. Then we can get the probability $P_{\mathrm{gc}}$ that $R_{\mathrm{min,\,2Gyr}} < 1$ kpc and $P_{\mathrm{MW}}$ that $R_{\mathrm{min,\,5Gyr}} < 25$ kpc \citep{Xu15}. They measure the probability that stars are derived from the Galactic center and the classfied criteria are shown in Table 1. Here, ``HVS'' represent the fastest stars in the Galaxy which are hypervelocity stars (HVS); ``OUT'' represent fast halo stars from the tidal debris of dwarf galaxy; ``HRS'' represent hyper-runaway star candidates; ``RS'' represent the runaway stars. 

As seen in Figure \ref{figure3}, the left two panel represent ``OUT'' candidate which is from the tidal debris of dwarf galaxy and the right two panel represent runaway stars and hyper-runaway stars which are thought to have formed in the disk and ejected into the halo.

\section{Conclusions and summary}

Based on the second Gaia data combined with observations from ground-based spectroscopic survey LAMOST DR5,  we cross-matched the initial sample and defined our HiVel star sample as those stars with $v_{\mathrm{gc}} > 0.85 v_{\mathrm{esc}}$, and derived final sample of 24 HiVels with reliable astrometric parameters and radial velocities. We studied the metallicity and [$\alpha$/Fe] distribution of our HiVel stars. While most of the HiVel stars are metal-poor and $\alpha$-enhanced. It shows some HiVel stars could have originated from the Galactic center or disk, while some from dwarf galaxies. To further understand the origin of HiVel stars, we traced the backwards orbits of each star in the Galactic potential to derive probability parameters which are used to classify these HiVel stars. According to the classified criteria,  5 stars are from the tidal debris of accreted and disrupted dwarf galaxy, 19 stars are runaway stars candidates which originate from the disk of the Galaxy and 6 of them are HRS candidates. There are two stars with high metallicity and low [$\alpha$/Fe]. One of them are ``OUT'' stars, which could from the dwarf galaxy. The other is ``RS'' star, it is similar to thick disk star according to its orbit.

\section*{Acknowledgements}

\par  We thank especially the referee for insightful comments and suggestions, which have improved the paper significantly. 
This work was supported by joint funding for Astronomy by the National Natural Science Foundation of China and the Chinese Academy of Science, under Grants U1231113.  This work was also by supported by the Special funds of cooperation between the Institute and the University of the Chinese Academy of Sciences, and China Scholarship Council (CSC).  In addition, this work was supported by the National Natural Foundation of China (NSFC No.11625313 and No.11573035).
HJN acknowledges funding from NSF grant AST 16-15688. Funding for SDSS-III has been provided by the Alfred P. Sloan Foundation, the Participating Institutions, the National Science Foundation, and the U.S. Department of Energy Office of Science. 
This project was developed in part at the 2016 NYC Gaia Sprint, hosted by the Center for Computational Astrophysics at the Simons Foundation in New York City. The Guoshoujing Telescope (the Large Sky Area Multi-Object Fiber Spectroscopic Telescope, LAMOST) is a National Major Scientific Project built by the Chinese Academy of Sciences. Funding for the project has been provided by the National Development and Reform Commission. LAMOST is operated and managed by the National Astronomical Observatories, Chinese Academy of Sciences. This work has made use of data from the European Space Agency (ESA) mission Gaia (http://www.cosmos.esa.int/gaia), processed by the Gaia Data Processing and Analysis Consortium (DPAC, http://www.cosmos.esa.int/web/gaia/dpac/consortium). Funding for DPAC has been provided by national institutions, in particular the institutions participating in the Gaia Multilateral Agreement.

\appendix  
\renewcommand{\appendixname}{Appendix~\Alph{section}}  

\begin{center}
	\renewcommand\arraystretch{1.4}
	\begin{longtable}{rrrrrrrr}
	\caption{Atmospheric parameters and positions for 24 HiVel stars which is classified as three subclass}\\
\hline
\hline
Notation &            source-id &         ra &       dec &  T$_{\mathrm{eff}}$ &  log(g) &            [Fe/H] &    [$\alpha$/Fe] \\
 &             &         (deg) &       (deg) &  (K) &  (dex) &            (dex) &    (dex) \\
\hline\\
\textbf{OUT candidates} &&&&&&&\\
GLHV-9 &  4554190291969378048 &  263.38934 &  20.32675 &                4502 &    0.95 &  -2.12 $\pm$ 0.02 &                - \\
GLHV-14 &  1002667880552099328 &  106.79127 &  59.57526 &                5309 &    3.81 &  -0.07 $\pm$ 0.03 &   0.05 $\pm$ 0.1 \\
GLHV-19 &  3905884598043829504 &  181.25391 &   9.45570 &                5062 &    2.47 &  -1.59 $\pm$ 0.01 &  0.22 $\pm$ 0.03 \\
GLHV-20 &  4586965565362654464 &  278.93102 &  27.96378 &                4437 &    0.78 &  -2.07 $\pm$ 0.07 &  0.29 $\pm$ 0.04 \\
GLHV-23 &  4450458649852400640 &  242.69749 &   7.16000 &                4752 &    1.70 &   -1.6 $\pm$ 0.02 &  0.26 $\pm$ 0.03 \\
\textbf{HRS candidates} &&&&&&&\\
GLHV-12 &  4443717204762199552 &  256.21663 &   9.35535 &                4389 &    1.26 &  -1.02 $\pm$ 0.04 &  0.17 $\pm$ 0.03 \\
GLHV-13 &  4015088951907615744 &  185.47443 &  31.07946 &                5765 &    3.39 &  -1.02 $\pm$ 0.02 &  0.22 $\pm$ 0.03 \\
GLHV-16 &  4491237203962217088 &  259.88425 &   8.67324 &                4710 &    1.53 &  -1.75 $\pm$ 0.03 &                - \\
GLHV-21 &  4443836776652403072 &  257.86499 &  10.21303 &                4469 &    1.31 &  -1.19 $\pm$ 0.03 &  0.24 $\pm$ 0.03 \\
GLHV-22 &  2629296824480015744 &  335.83338 &  -2.51967 &                5212 &    3.18 &  -1.02 $\pm$ 0.01 &  0.19 $\pm$ 0.04 \\
GLHV-24 &  1383279090527227264 &  240.33735 &  41.16677 &                4803 &    1.89 &   -1.4 $\pm$ 0.02 &  0.28 $\pm$ 0.03 \\
\textbf{RS candidates} &&&&&&&\\
GLHV-1 &  2503491051919554304 &   39.87357 &   3.10509 &                6236 &    4.16 &  -1.37 $\pm$ 0.02 &  0.47 $\pm$ 0.09 \\
GLHV-2 &  1203885900077833984 &  237.11051 &  19.28888 &                4533 &    1.04 &  -2.21 $\pm$ 0.02 &  0.26 $\pm$ 0.03 \\
GLHV-3 &  4485842140925368832 &  265.28990 &   6.23972 &                4149 &    0.62 &  -1.21 $\pm$ 0.05 &                - \\
GLHV-4 &  2106519830479009920 &  285.48442 &  45.97166 &                4468 &    1.33 &  -1.15 $\pm$ 0.02 &   0.2 $\pm$ 0.03 \\
GLHV-5 &  1268023196461923712 &  225.78358 &  26.24632 &                4925 &    2.13 &  -1.53 $\pm$ 0.01 &  0.22 $\pm$ 0.03 \\
GLHV-6 &  3784964943489710592 &  169.35630 &  -5.81538 &                4851 &    2.08 &  -1.14 $\pm$ 0.09 &                - \\
GLHV-7 &  1597988246569491968 &  233.62013 &  54.30564 &                5448 &    2.82 &  -1.05 $\pm$ 0.04 &  0.25 $\pm$ 0.04 \\
GLHV-8 &   598766750854551168 &  131.80999 &  11.03167 &                6311 &    4.18 &  -0.24 $\pm$ 0.01 &   0.06 $\pm$ 0.1 \\
GLHV-10 &  1255095276181144320 &  218.71274 &  25.16609 &                5439 &    4.50 &  -1.24 $\pm$ 0.05 &  0.32 $\pm$ 0.09 \\
GLHV-11 &   330414789019026944 &   29.21933 &  36.66581 &                4976 &    2.15 &  -2.03 $\pm$ 0.01 &  0.19 $\pm$ 0.03 \\
GLHV-15 &  1341901032000157056 &  258.13448 &  40.47352 &                4660 &    1.40 &  -2.04 $\pm$ 0.05 &                - \\
GLHV-17 &  1552278116525348096 &  204.66905 &  48.15653 &                5691 &    3.88 &  -0.85 $\pm$ 0.02 &   0.14 $\pm$ 0.1 \\
GLHV-18 &  3736372993468775424 &  197.96401 &  11.28944 &                4946 &    2.21 &  -1.38 $\pm$ 0.04 &  0.21 $\pm$ 0.03 \\
\hline
\end{longtable}
\end{center}

\begin{center}
	\renewcommand\arraystretch{1.4}
	\begin{longtable}{rrrrrrrrr}
		\caption{Velocities and distances for 24 HiVel stars which is classified as three subclass}\\
		\hline
		\hline
Notation &         $rv_L$ &        $rv_G$ &   $\overline{rv}$ &                   $d$ &              $r_{gc}$ &             $v_{gc}$ &  $P_{\mathrm{MW}}$ &  $P_{\mathrm{ub}}$ \\
 &         (km/s) &        (km/s) &        (km/s) &                    kpc &              kpc &             (km/s) &   &   \\
		\hline\\
		\textbf{OUT candidates} &&&&&&&&\\
  GLHV-9 &   -315 $\pm$ 6 &  -319 $\pm$ 1 &    -319 $\pm$ 1 &  $10.3^{+2.0}_{-1.9}$ &   $7.9^{+1.3}_{-1.0}$ &  $554^{+134}_{-124}$ &               0.42 &               0.58 \\
GLHV-14 &    -65 $\pm$ 4 &   -64 $\pm$ 1 &     -64 $\pm$ 1 &   $2.1^{+0.8}_{-0.6}$ &  $10.1^{+0.7}_{-0.5}$ &   $583^{+129}_{-93}$ &               0.21 &               0.79 \\
GLHV-19 &    148 $\pm$ 5 &   149 $\pm$ 1 &     149 $\pm$ 1 &   $2.5^{+0.2}_{-0.2}$ &   $8.6^{+0.1}_{-0.1}$ &    $519^{+69}_{-61}$ &               0.48 &               0.52 \\
GLHV-20 &  -238 $\pm$ 10 &  -234 $\pm$ 4 &    -235 $\pm$ 4 &  $10.5^{+1.6}_{-1.4}$ &   $9.3^{+1.1}_{-0.9}$ &    $547^{+93}_{-85}$ &               0.36 &               0.64 \\
GLHV-23 &   -120 $\pm$ 6 &  -120 $\pm$ 1 &    -120 $\pm$ 1 &   $7.5^{+1.8}_{-1.2}$ &   $5.8^{+0.7}_{-0.2}$ &  $711^{+194}_{-127}$ &               0.08 &               0.92 \\
\textbf{HRS candidates} &&&&&&&&\\
GLHV-12 &   -276 $\pm$ 4 &  -277 $\pm$ 0 &    -277 $\pm$ 0 &   $7.2^{+1.6}_{-1.1}$ &   $5.3^{+0.5}_{-0.1}$ &  $593^{+163}_{-102}$ &               1.00 &               0.62 \\
GLHV-13 &    -56 $\pm$ 8 &   -55 $\pm$ 1 &     -55 $\pm$ 1 &   $2.1^{+0.4}_{-0.3}$ &   $8.7^{+0.1}_{-0.1}$ &    $516^{+99}_{-72}$ &               0.60 &               0.50 \\
GLHV-16 &   -302 $\pm$ 6 &  -296 $\pm$ 2 &    -297 $\pm$ 1 &   $7.0^{+1.5}_{-1.2}$ &   $5.1^{+0.3}_{-0.1}$ &  $603^{+149}_{-115}$ &               1.00 &               0.63 \\
GLHV-21 &   -184 $\pm$ 4 &  -182 $\pm$ 1 &    -183 $\pm$ 1 &  $13.0^{+2.8}_{-2.1}$ &   $8.5^{+2.3}_{-1.5}$ &   $619^{+120}_{-90}$ &               1.00 &               0.84 \\
GLHV-22 &   -215 $\pm$ 5 &  -220 $\pm$ 4 &    -218 $\pm$ 3 &   $0.9^{+0.0}_{-0.0}$ &   $8.0^{+0.0}_{-0.0}$ &    $568^{+22}_{-22}$ &               1.00 &               0.98 \\
GLHV-24 &   -179 $\pm$ 5 &  -181 $\pm$ 2 &    -181 $\pm$ 2 &   $6.2^{+0.7}_{-0.5}$ &   $8.8^{+0.3}_{-0.2}$ &    $629^{+88}_{-66}$ &               1.00 &               0.96 \\
\textbf{RS candidates} &&&&&&&&\\
GLHV-1 &   361 $\pm$ 10 &   363 $\pm$ 1 &     363 $\pm$ 1 &   $0.4^{+0.0}_{-0.0}$ &   $8.5^{+0.0}_{-0.0}$ &      $444^{+4}_{-4}$ &               1.00 &               0.00 \\
GLHV-2 &   -255 $\pm$ 6 &  -246 $\pm$ 2 &    -246 $\pm$ 2 &   $7.6^{+1.4}_{-1.3}$ &   $7.4^{+0.7}_{-0.4}$ &    $479^{+42}_{-37}$ &               0.81 &               0.19 \\
GLHV-3 &    -17 $\pm$ 4 &    -8 $\pm$ 1 &      -8 $\pm$ 1 &   $6.9^{+2.3}_{-1.3}$ &   $4.9^{+0.5}_{-0.1}$ &   $563^{+129}_{-76}$ &               1.00 &               0.48 \\
GLHV-4 &   -215 $\pm$ 4 &  -213 $\pm$ 1 &    -213 $\pm$ 1 &   $6.4^{+0.8}_{-0.6}$ &   $9.2^{+0.4}_{-0.3}$ &    $457^{+51}_{-38}$ &               1.00 &               0.17 \\
GLHV-5 &   -275 $\pm$ 5 &  -277 $\pm$ 2 &    -277 $\pm$ 2 &   $3.9^{+0.4}_{-0.3}$ &   $7.6^{+0.0}_{-0.0}$ &    $474^{+56}_{-42}$ &               1.00 &               0.18 \\
GLHV-6 &    128 $\pm$ 7 &   126 $\pm$ 1 &     126 $\pm$ 1 &   $3.3^{+0.5}_{-0.3}$ &   $9.0^{+0.2}_{-0.1}$ &    $457^{+55}_{-36}$ &               0.97 &               0.18 \\
GLHV-7 &     41 $\pm$ 8 &    40 $\pm$ 1 &      40 $\pm$ 1 &   $1.2^{+0.1}_{-0.0}$ &   $8.2^{+0.0}_{-0.0}$ &    $464^{+23}_{-20}$ &               1.00 &               0.01 \\
GLHV-8 &     75 $\pm$ 5 &    72 $\pm$ 2 &      72 $\pm$ 2 &   $3.2^{+0.7}_{-0.5}$ &  $10.7^{+0.6}_{-0.4}$ &    $479^{+64}_{-48}$ &               0.61 &               0.39 \\
GLHV-10 &    367 $\pm$ 8 &   367 $\pm$ 2 &     367 $\pm$ 1 &   $0.2^{+0.0}_{-0.0}$ &   $8.1^{+0.0}_{-0.0}$ &      $470^{+1}_{-1}$ &               1.00 &               0.00 \\
GLHV-11 &   -127 $\pm$ 5 &  -121 $\pm$ 2 &    -122 $\pm$ 2 &   $1.9^{+0.2}_{-0.1}$ &   $9.6^{+0.1}_{-0.1}$ &    $491^{+48}_{-41}$ &               1.00 &               0.37 \\
GLHV-15 &  -219 $\pm$ 11 &  -222 $\pm$ 2 &    -222 $\pm$ 2 &   $6.1^{+0.9}_{-0.6}$ &   $8.4^{+0.4}_{-0.2}$ &    $509^{+88}_{-60}$ &               1.00 &               0.47 \\
GLHV-17 &    -85 $\pm$ 6 &   -84 $\pm$ 2 &     -84 $\pm$ 2 &   $2.1^{+0.2}_{-0.1}$ &   $8.6^{+0.1}_{-0.0}$ &    $490^{+53}_{-38}$ &               1.00 &               0.31 \\
GLHV-18 &    380 $\pm$ 7 &   378 $\pm$ 1 &     378 $\pm$ 1 &   $3.2^{+0.3}_{-0.3}$ &   $8.1^{+0.1}_{-0.0}$ &    $499^{+21}_{-15}$ &               1.00 &               0.18 \\
		\hline
	\end{longtable}
\end{center}

\end{document}